\documentclass[11pt,reqno,twoside]{article}
\usepackage{amsmath,amssymb,theorem,color,epsfig,epic,subfigure,amsfonts,hhline,graphicx}

\theorembodyfont{\itshape}
\newtheorem{thm}{Theorem}[section]
\newtheorem{lem}[thm]{Lemma}
\newtheorem{prop}[thm]{Proposition}
{\theorembodyfont{\upshape}

\newtheorem{rem}[thm]{Remark}
\newtheorem{ex}[thm]{Example}
}
\newtheorem{cor}[thm]{Corollary}

\newcommand{\Proof}[1][]{\noindent{\itshape Proof#1. }}
\newcommand{\EndProof}{\hfill$\Box$\bigskip}

\makeatletter
    
    \@addtoreset{equation}{section}
\makeatother

\def\hat{\widehat}
\def\tilde{\widetilde}

\def\ol{\overline}

\def\Tr{\text{Tr}}

\def\d{\delta}    
\def\a{\alpha}    \def\b{\beta}              \def\d{\delta}
\def\D{\Delta}    \def\e{\varepsilon}        
             
\def\l{\lambda}   \def\L{\Lambda}             
\def\r{\rho}               
\def\p{\pi}              \def\s{\sigma}

\begin{document}
\title{Central limit theorems for open quantum random walks on the crystal lattices}
\author{Chul Ki Ko\footnote{University College, Yonsei University, 85 Songdogwahak-ro,
Yeonsu-gu, Incheon 21983, Korea. E-mail: kochulki@yonsei.ac.kr}, Norio Konno\footnote{Department of Applied Mathematics, Faculty of Engineering, Yokohama National University, Hodogaya, Yokohama 240-8501, Japan. E-mail: konno@ynu.ac.jp}, Etsuo Segawa\footnote{Graduate School of Information Sciences, Tohoku University, 6-3-09 Aramaki Aza, Aoba, Sendai, Miyagi, 980-8579, Japan. E-mail: e-segawa@m.tohoku.ac.jp}, and
   Hyun Jae Yoo\footnote{Department of Applied Mathematics,
Hankyong National University, 327 Jungang-ro, Anseong-si,
Gyeonggi-do 17579, Korea. E-mail: yoohj@hknu.ac.kr} \footnote{Corresponding author}  }
\date{ }
   \maketitle

\begin{abstract}
We consider the open quantum random walks on the crystal lattices and investigate the central limit theorems for the walks. On the integer lattices the open quantum random walks satisfy the central limit theorems as was shown by Attal, {\it et al}. In this paper we prove the central limit theorems for the open quantum random walks on the crystal lattices. We then provide with some examples for the Hexagonal lattices. We also develop the Fourier analysis on the crystal lattices. This leads to construct the so called dual processes for the open quantum random walks. It amounts to get Fourier transform of the probability densities, and it is very useful when we compute the characteristic functions of the walks. In this paper we construct the dual processes for the open quantum random walks on the crystal lattices providing with some examples.
\end{abstract}

\section{Introduction}\label{sec:introduction} 

The purpose of this paper is to construct the open quantum random walks on the crystal lattices and investigate the asymptotic behavior, namely central limit theorems.

The unitary quantum walks have been developed and applied as a tool for quantum algorithms, and it succeeded by its power of speeding up in certain search algorithms \cite{Am, Ch, CCDFGS, Sz}. Since it was mathematically formulated many properties of quantum walks have been known, especially the asymptotic behavior of the quantum walks was shown \cite{ABNVW, GJS, Ke, K1, K2}. More precisely, it was proved that the quantum walks, when they are scaled by $1/n$, have limit distributions with certain densities, which are drastically different from Gaussian, the limit distributions of the classical randoms walks resulting from the central limit theorem \cite{GJS, K1, K2}. 

Recently, a new type of quantum walks, so called open quantum random walks (OQRWs hereafter) was introduced \cite{AG-PS, APSS, APS}. The OQRWs were developed to formulate the dissipative quantum computing algorithms and dissipative quantum state preparation \cite{APS}. The decoherence and dissipation occur by the interaction of a system with environment and one needs to develop a proper quantum walk so that the dissipativity can be implemented. The works of \cite{AG-PS, APSS, APS} aim to fulfill this requirement. The OQRWs are not unitary evolutions of states, contrary to the early developed unitary quantum walks (it was thus named). By the procedure of quantum trajectories, which amounts to a repeated measurement of the particle at each step and an application of a completely positive map, the OQRWs are simulated by Markov chains on the product of position and state spaces \cite{AG-PS, APSS, APS} (see section \ref{sec:oqrw_on_crystal_lattice} for the details).  

In the paper \cite{AG-PS}, Attal {\it et al.} proved the central limit theorems for OQRWs on the integer lattice $\mathbb Z^d$. This result typically shows that the behavior of OQRWs and unitary quantum walks are much different. On the other hand, when we consider the dynamics on the integer lattices, we can develop Fourier transforms. In \cite{KY}, Konno and Yoo developed the Fourier transform theory for the OQRWs on the integer lattices, and by it the so called dual process was constructed. It is in a sense the process of Fourier transforms of probability distributions. Some related works on the central limit theorems for OQRWs, one can find in the references \cite{Br, SP}.

In this paper we construct OQRWs on the crystal lattices. The crystal lattices are the structures which have regularity globally, like integer lattices, but may have further structure locally (see subsection \ref{subsec:crystal_lattices} for the definition). Therefore, not only the integer lattices belong to this class but more fruitful structures can be considered. The goal of the paper is two-fold: one is to show the central limit theorems for the OQRWs on the crystal lattices and the other is to construct the dual processes by using a Fourier transform theory on the crystal lattices. Following the superb method developed in \cite{AG-PS} we could show the central limit theorems. We will provide with some examples for the Hexagonal lattices. We then develop a Fourier transform theory and construct the dual processes as was done in \cite{KY}. By revisiting the examples we will see that the central limit theorems can be also obtained by the dual processes. In some examples it even provides a better understanding of the dynamics. We remark that recently the present authors considered the orbits, or the support  of the scaled unitary quantum walks on the crystal lattices \cite{KKSY}.

This paper is organized as follows. In section \ref{sec:oqrw_on_crystal_lattice} we introduce the crystal lattices and construct OQRWs on them. In section \ref{sec:clt}, we show the central limit theorems (Theorem \ref{thm:clt}). Section \ref{sec:examples} is devoted to the examples. Typically we will consider Hexagonal lattices. We give two examples which have nonzero- and zero-covariances, respectively, in the limit. In section \ref{sec:dual_process}, we construct dual processes after a short introduction of Fourier analysis on the crystal lattices. The examples mentioned above are revisited for comparison. Appendix A gives a proof for the central limit theorem. We follow the methods in \cite{AG-PS} with a suitable modification. In the Appendix B and C, we provide with analytic proofs for some technical results that are used in the examples.

\section{OQRWs on the crystal lattices}\label{sec:oqrw_on_crystal_lattice}
\subsection{Crystal lattices}\label{subsec:crystal_lattices}
In this subsection we introduce the crystal lattices as was done in \cite{KKSY}. Let $G_0=(V_0,E_0)$ be a finite graph which may have multi edges and self loops. 
We use the notation $A(G_0)$ for the set of symmetric arcs induced by $E_0$. 
The homology group of $G_0$ with integer coefficients is denoted by $H_1(G_0,\mathbb{Z})$. 
The abstract periodic lattice $\mathbb L$ induced by a subgroup $H\subset H_1(G_0,\mathbb{Z})$ is denoted by $H_1(G_0,\mathbb{Z})/H$~\cite{Su}. 

Let the set of basis of $H_1(G_0,\mathbb{Z})$ be $\{C_1,C_2,\dots,C_{b_1}\}$ corresponding to fundamental cycles of $G_0$, 
where $b_1$ is the first Betti number of $G_0$. 
The spanning tree induced by $\{C_1,C_2,\dots,C_{b_1}\}$ is denoted by $\mathbb{T}_0$. 
We can take a one-to-one correspondence between $\{C_1,C_2,\dots,C_{b_1}\}$ and $A(\mathbb{T}_0)^{c}$; 
we describe $C(e)\in \{C_1,C_2,\dots,C_{b_1}\}$ as the fundamental cycle corresponding to $e\in A(\mathbb{T}_0)^{c}$ 
so that $C(e)$ is the cycle generated by adding $e$ to $\mathbb{T}_0$. 
Let $d$ be the number of generators of the quotient group $H_1(G_0,\mathbb{Z})/H$. By taking a set of generating vectors $\{\hat \theta(e):e\in A(\mathbb{T}_0)^{c}\}$ (we suppose $\hat{\theta}(\bar{e})=-\hat{\theta}(e)$, where $\bar{e}$ means the reversed arc of $e$), we may consider $\mathbb L$ as a subset of $\mathbb R^d$ isomorphic to $\mathbb Z^d$ . In other words, we may think 
\[
\mathbb L=\left\{\sum n_e\hat\theta(e):e\in A(\mathbb{T}_0)^{c},\,\,n_e\in \mathbb Z\right\}.
\]

Let us define a covering graph $G=(V,A)$ of $G_0$ by the lattice $\mathbb L$. For it, define  
$\phi: A(\mathbb{T}_0) \to \mathbb{R}^d$ so that 
  $ \phi(\bar{e})=-\phi(e)$ for every $e\in A_0$.  
We also define $\phi_0: V_0\to \mathbb{R}^d$ so that 
	$ \phi(e)=\phi_0(\mathrm t(e))-\phi_0(\mathrm o(e)) $
for every $e\in A(\mathbb{T}_0)$ by fixing a point $\phi_0(v_0)$ at some vertex $v_0\in V_0$. Here $\mathrm t(e)$ and $\mathrm o(e)$ denote the terminal and origin of the arc $e$, respectively. Now the covering graph $G=(V,A)$ is defined as follows.
\begin{eqnarray*}
        V &=& \mathbb L+\phi_0(V_0) \cong \mathbb L\times \phi_0(V_0); \\
        A&= & \cup_{x\in \mathbb L} \left\{ \left((x,\mathrm o(e)),(x,\mathrm t(e))\right) \;|\; e\in A(\mathbb{T}_0) \right\}  \\
        	&& \quad \cup \left(\cup_{x\in \mathbb L} \left\{ \left((x,\mathrm o(e)),(x+\hat\theta(e),\mathrm t(e))\right) \;|\; e\in A^{c}(\mathbb{T}_0) \right\}\right).
        \end{eqnarray*}
The covering graph $G=(V,A)$ is called a crystal lattice.        

\begin{figure}[h]
\begin{center}
\includegraphics[width=0.25\textwidth]{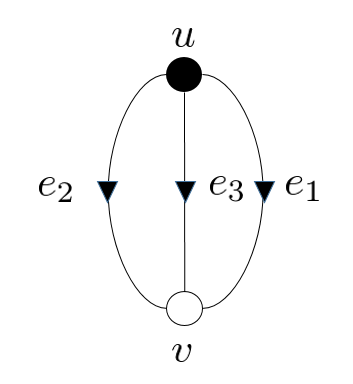}~~~~~~~~\includegraphics[width=0.5\textwidth]{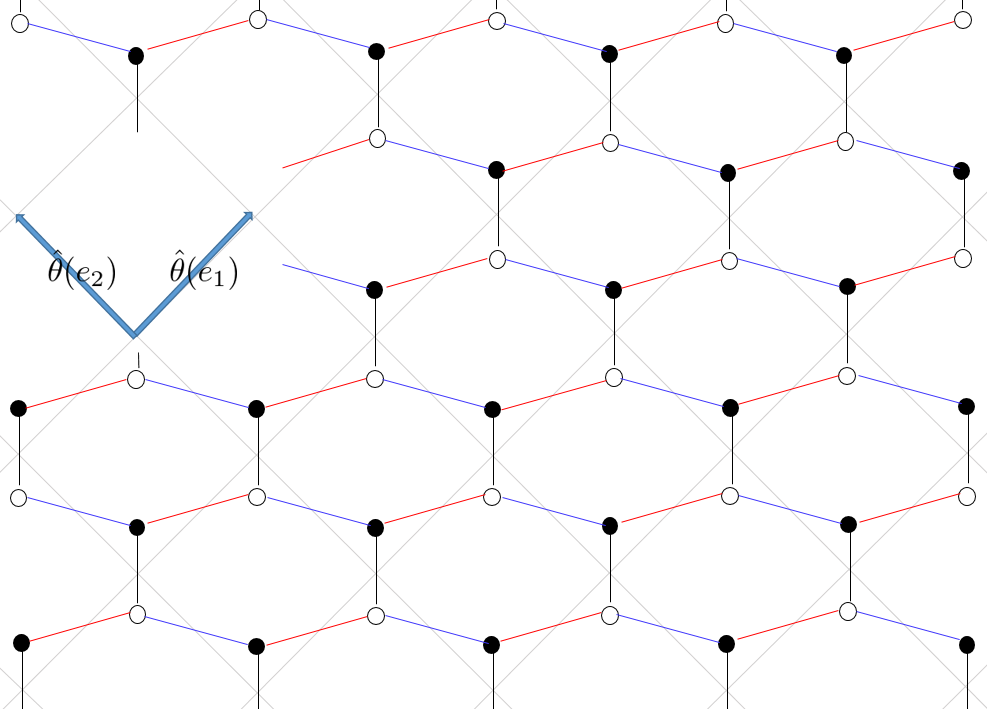}
\caption{Hexagonal lattice: underlying graph $G_0$ for Hexagonal lattice (left), Hexagonal lattice (right)}\label{fig:hexagonal_lattice}
\end{center}
\end{figure}


We take $\hat\theta(e)\equiv 0$ for $e\in A(\mathbb{T}_0)$ and choose $e_{i_1},\dots,e_{i_d}$ from $A(\mathbb{T}_0)^c$ so that 
$\hat{\theta}_1:=\hat{\theta}(e_{i_1}),\dots,\hat{\theta}_d:=\hat{\theta}(e_{i_d})$ span $\mathbb{R}^d$. We further suppose that for all $e\in A(G_0)$, $\hat\theta(e)\in \{\sum_{i=1}^dn_i\hat\theta_i:n_i\in \mathbb Z,\,i=1,\cdots,d\}$, and for any two arcs $e_i$ and $e_j$ in $A(\mathbb T_0)^c$, $\hat\theta(e_i)$ and $\hat\theta(e_j)$ are linearly independent unless $e_j=\ol{e}_i$.
We define a $d\times d$ matrix by
	\begin{equation}\label{eq:transformation_matrix} \Theta:= ([\hat{\theta}_1,\dots,\hat{\theta}_{d}]^{-1})^T. 
	\end{equation}
Notice that if $\{{\bf e}_i:i=1,\cdots,d\}$ is the canonical basis for $\mathbb R^d$, then we have 
\begin{equation}\label{eq:transformation}
{\bf e}_i=\sum_{j=1}^d\Theta_{ij}\hat\theta_j.
\end{equation}
The matrix $\Theta$ will take a crucial role when we consider Fourier transforms.	

\subsection{OQRWs on the crystal lattices}

We let $\mathcal K:=l^2(\mathbb L)$  and by $\{|x\rangle:x\in \mathbb L\}$ we denote the canonical orthonormal basis of $\mathcal K$. Let $\mathcal H$ be a finite dimensional Hilbert space and for each $u\in V_0$, let $\mathcal H_u$ be a copy of $\mathcal H$. Define 
\[
\mathfrak h:=\oplus_{u\in V_0}\mathcal H_u.
\]
$\mathfrak h$ represents an intrinsic structure at each site of $\mathbb L$. The Hilbert space $\mathfrak h\otimes \mathcal K$ is the base Hilbert space on which our OQRWs are working. For each $e\in A(G_0)$, $e=(u,v)$, we let $B(e)$ be a bounded linear operator from $\mathcal H_u$ to $\mathcal H_v$ satisfying 
\begin{equation}\label{eq:Kraus_point}
\sum_{\substack{e\in A(G_0);\\ \mathrm{o}(e)=u}}B^*(e)B(e)=I_{\mathcal H_u}\quad \text{for all }u\in V_0.
\end{equation}
Whenever there is no danger of confusion we also understand $\mathcal H_u$ as a subspace of $\mathfrak h$. With this convention, $B(e)$ (using the same symbol by abuse of notations) is a bounded linear operator on $\mathfrak h$ and satisfies
\begin{equation}\label{eq:Kraus_local}
\sum_{e\in A(G_0)}B^*(e)B(e)=\sum_{u\in V_0}\sum_{\substack{e\in A(G_0);\\\mathrm{o}(e)=u}}B^*(e)B(e)=\sum_{u\in V_0}I_{\mathcal H_u}=I_\mathfrak h.
\end{equation}
The operators $\{B(e):e\in A(G_0)\}$ will constitute the Kraus representation of our OQRWs on the crystal lattices. For that we define for each $x\in \mathbb L$ and $e\in A(G_0)$, a bounded linear operator $L_x^e$ on $\mathfrak h\otimes \mathcal K$ by 
\begin{equation}\label{eq:Kraus_operators} 
L_x^e:=B(e)\otimes|x+\hat\theta(e)\rangle\langle x|.
\end{equation}
We can check the following property.
\begin{lem}\label{lem:Kraus}
\begin{equation}\label{eq:Kraus_global}
\sum_{x\in \mathbb L}\sum_{e\in A(G_0)}\left(L_x^e\right)^*L_x^e=I_{\mathfrak h\otimes \mathcal K}.
\end{equation}
\end{lem} 
\Proof
By \eqref{eq:Kraus_local},
\begin{eqnarray*} 
\sum_{x\in \mathbb L}\sum_{e\in A(G_0)}\left(L_x^e\right)^*L_x^e&=&\sum_{x\in \mathbb L}\sum_{e\in A(G_0)}B(e)^*B(e)\otimes |x\rangle\langle x|\\
&=&\sum_{x\in \mathbb L} I_{\mathfrak h}\otimes |x\rangle\langle x|\\
&=&I_{\mathfrak h\otimes \mathcal K}. 
\end{eqnarray*}
\EndProof\\
The OQRW is a completely positive linear operator on the ideal $\mathcal I_1$ of trace class operators on ${\mathfrak h\otimes \mathcal K}$ defined by 
\begin{equation}\label{eq:oqrw}
\mathcal M(\rho):=\sum_{x\in \mathbb L}\sum_{e\in A(G_0)}L_{x}^{e}\rho (L_{x}^{e})^*.
\end{equation}
Let us consider a special class of states (density operators) on $\mathfrak h\otimes \mathcal K$ of the form 
\begin{equation}\label{eq:states}
\rho=\sum_{x\in  \mathbb L }\left(\oplus_{u\in V_0}\rho_{(x,u)}\right)\otimes|x\rangle\langle x|.
\end{equation}
Here, for each pair $(x,u)\in \mathbb L\times V_0$, $\rho_{(x,u)}$ is a positive definite operator on $\mathcal H_u$ and satisfies 
\[
\sum_{x\in \mathbb L}\sum_{u\in V_0}\Tr(\rho_{(x,u)})=1.
\]
The value $\sum_{u\in V_0}\Tr(\rho_{(x,u)})$ is understood as a probability of finding the particle at site $x\in \mathbb L$ when the state is $\rho$. We check that if the state has the form in \eqref{eq:states}, $\rho=\sum_{x\in  \mathbb L }\left(\oplus_{u\in V_0}\rho_{(x,u)}\right)\otimes|x\rangle\langle x|$, $\mathcal M(\rho)$ has the form
\begin{equation}\label{eq:oqrw_form}
\mathcal M(\rho)= \sum_{x\in  \mathbb L }\left(\oplus_{u\in V_0}\rho'_{(x,u)}\right)\otimes|x\rangle\langle x|,
\end{equation}
where
\[
\rho'_{(x,u)}=\sum_{ \substack{e\in A(G_0);\\\mathrm{t}(e)=u}}B(e)\rho_{(x-\hat\theta(e),\mathrm{o}(e))}B(e)^*.
\]
From now on we assume that $\mathcal M$ is defined on the set of states of the form in \eqref{eq:states}.

Let $X$ denote the random variable representing the position of the particle, or the walker. Starting from the initial state $\rho$ in \eqref{eq:states}, the probability of finding the particle at site $x\in \mathbb L$ after a one-step evolution is given by 
\[
\mathbb{P}(X=x)=\sum_{u\in V_0}\Tr(\rho'_{(x,u)}).
\]
As was introduced in \cite{AG-PS,APSS}, let $(\rho_n,X_n)_{n\ge 0}$ denote the Markov chain of quantum trajectory procedure. This is obtained by repeatedly applying the completely positive map $\mathcal M$ and a measurement of the position on $\mathcal K$. More precisely, denoting $\mathcal E(\mathfrak h)$ the space of states on $\mathfrak h$,   $(\rho_n,X_n)_{n\ge 0}$ is a Markov chain on the state space $\mathcal E(\mathfrak h)\times \mathbb L$ for which the transition rule is defined as follows: from a point $(\rho,x)\in \mathcal E(\mathfrak h)\times \mathbb L$ it jumps to the point
\[
\left(\frac1{p(e)}B(e)\rho B(e)^*, x+\hat\theta(e)\right)\in \mathcal E(\mathfrak h)\times \mathbb L,
\]
with probability 
\[
p(e)=\Tr(B(e)\rho B(e)^*).
\] 

\section{Central limit theorem}\label{sec:clt}

In this section we discuss the central limit theorem for the OQRWs on the crystal lattices. The same study for the OQRWs on the integer lattices $\mathbb Z^d$ was done in \cite{AG-PS}. Here we follow the same stream lines of \cite{AG-PS} with slight modifications.

\subsection{Preparation}\label{subsec:preparation}

We let 
\begin{equation}\label{eq:cp_equation}
\mathcal L(\rho):=\sum_{e\in A(G_0)}B(e)\rho B(e)^*,\quad \rho\in \mathcal E(\mathfrak h).
\end{equation}
We assume the following hypothesis.
\begin{enumerate}
\item[(H)] $\mathcal L$ admits a unique invariant state $\rho_\infty$.
\end{enumerate}
\begin{rem}\label{rem:ergodicity}
The existence of an invariant state to the equation \eqref{eq:cp_equation} follows from an ergodic theorem \cite{KM}. In fact, for any initial state $\rho_0$, the time average
\[
\frac1n\sum_{k=0}^{n-1}\mathcal L^k(\rho_0)
\]
converges almost surely (in a suitable probability space) to an invariant state $\rho_\infty\in \mathcal E(\mathfrak h)$ (see also \cite{AG-PS}).
\end{rem} 
Let us define 
\begin{equation}\label{eq:mean}
m:=\sum_{e\in A(G_0)}\Tr\left(B(e)\rho_\infty B(e)^*\right)\hat\theta(e).
\end{equation}
\begin{lem}\label{lem:operator_solution}
For any $l\in \mathbb R^d$, the equation
\begin{equation}\label{eq:operator_equation}
L-{\mathcal L}^*(L)=\sum_{ e\in A(G_0) }B(e)^*B(e)\left(\hat\theta(e)\cdot l\right)-(m\cdot l)I
\end{equation}
admits a solution. The difference between any two solutions of \eqref{eq:operator_equation} is a multiple of the identity.
\end{lem}
\Proof
By \eqref{eq:mean} we have for any $l\in \mathbb R^d$,
\[
\sum_{ e\in A(G_0) }\Tr\left(B(e)\rho_\infty B(e)^*\right)\hat\theta(e)\cdot l=m\cdot l.
\]
Hence 
\[
\Tr\Big(\rho_\infty\Big(\sum_{ e\in A(G_0) }B (e)^*B(e)\hat\theta(e)\cdot l-\big(m\cdot l\big)I\Big)\Big)=0.
\]
Thus
\[
\sum_{ e\in A(G_0) }B(e)^*B(e)\left(\hat\theta(e)\cdot l\right)-\big(m \cdot l\big)I\in \{\rho_\infty\}^\perp=\mathrm{Ran}(I-{\mathcal L}^*).
\]
The last equality follows from the fact that  
\[
\{\rho_\infty\}^\perp=\mathrm{Ker}(I-\mathcal L )^\perp=\ol{\mathrm{Ran}(I-{\mathcal L^* })}=\mathrm{Ran}(I-{\mathcal L}^*),
\] 
since $\mathfrak h$ is of finite dimensional. This proves the first part. The second part can be proven by the same argument that was used in \cite[Lemma 5.1]{AG-PS}.  
\EndProof\\
Let us denote the solution of \eqref{eq:operator_equation} corresponding to $l$ by $L_l $. In particular, for the basis vectors $\{\hat\theta_1,\cdots,\hat\theta_d\}$ of the lattice $\mathbb L$, we denote $L_i $ for $L_{\hat\theta_i} $, $i=1,\cdots,d$. Note that 
\begin{equation}\label{eq:expansion}
L_l =\sum_{i=1}^dl_iL_i,
\end{equation}
where $\{l_i\}$ are the coordinates of $l$ w.r.t. $\{\hat\theta_i\}$.

Recall the Markov chain $
(\rho_n,X_n)_{n\ge 0}$ on the state space $\mathcal{E}(\mathfrak h)\times\mathbb L$. We introduce a related Markov chain. The Markov chain $(\rho_n,Y_n)_{n\ge 0}$ is defined on the state space $\mathcal E(\mathfrak h)\times  A(G_0)$. The transition probabilities are given as follows. From the state $(\rho, e)$, it jumps to $(\rho',e')$ with probability $\Tr(B(e')\rho B(e')^*)$, where $\rho'=\frac1{\Tr(B(e')\rho B(e')^*)}(B(e')\rho B(e')^*)$. Notice that if we put $\D X_n:=X_n-X_{n-1}\in \{\hat\theta(e):\,e\in A(G_0)\}$, then $(\r_n,\D X_n)_{n\ge 0}$ is a Markov chain that is equivalent with  $(\rho_n,\hat\theta(Y_n))_{n\ge 0}$. The Markov operator (transition operator) for the Markov chain $(\rho_n,Y_n)_{n\ge 0}$ is denoted by $P$.
\begin{rem}\label{rem:jumps}
We emphasize here that if $(\rho,e)$ is the present state for the Markov chain $(Y_n)$ and particularly if $\rho$ is supported on $\mathcal H_u$ for some $u\in V_0$ (recall that $\mathfrak h=\oplus_{u\in V_0}\mathcal H_u$), then it jumps to some $(\rho',e')$ where $e'$ must satisfy $\mathrm{o}(e)=u$, since $B(e')\rho B(e')^*=0$ if $\mathrm{o}(e')\neq u$.
\end{rem}
Let us consider the Poisson equation \cite{AG-PS}:
\begin{equation}\label{eq:Poisson}
(I-P)f(\rho,e)=\hat\theta(e)\cdot l-m \cdot l.
\end{equation}
\begin{lem}\label{rlem:Poisson}
The equation \eqref{eq:Poisson} admits a solution which is 
\[
f(\rho,e)=\mathrm{Tr}(\rho L_l)+\hat\theta(e)\cdot l.
\]
\end{lem}
\Proof
For the function $f(\rho,e)$ in the statement, we have
\begin{eqnarray*}
(I-P)f(\rho,e)&=&\mathrm{Tr}(\rho L_l)+\hat\theta(e)\cdot l\\
&&-\sum_{ e'\in A(G_0) }\Big(\Tr\big(B(e')\rho B(e')^*L_l\big)+ \Tr\big(B(e')\rho B(e')^*\big)\hat\theta(e')\cdot l\Big)\\
&=&\Tr\Big(\rho\Big(L_l-{\mathcal{L}}^*(L_l)-\sum_{e'\in A(G_0)}B(e')^*B(e')\hat\theta(e')\cdot l\Big)\Big)+\hat\theta(e)\cdot l\\
&=&\hat\theta(e)\cdot l-m\cdot l.
\end{eqnarray*}
The proof is completed.
\EndProof

\subsection{Central limit theorem}\label{subsec:clt}

In this subsection we present the central limit theorem for the OQRWs on the crystal lattices. All the ingredients needed to show the central limit theorem are prepared in the previous subsection. The main result of this paper is the following theorem. 
\begin{thm}\label{thm:clt}
Consider the open quantum random walk on a crystal lattice (embedded in $\mathbb R^d$). Assume that the completely positive map
\[
\mathcal L(\rho)=\sum_{e\in A(G_0)}B(e)\rho B(e)^*
\]
admits a unique invariant state $\rho_\infty$ on $\mathfrak h$. Let $(\rho_n,X_n)_{n\ge 0}$ be the quantum trajectory process associated to this OQRW. Then,
\[
\frac{X_n-nm}{\sqrt{n}}
\]
converges in law to the Gaussian distribution $N(0,\Sigma)$ in $\mathbb R^d$, with covariance matrix $\Sigma=(C_{ij})_{i,j=1}^d$ given by
\begin{eqnarray}\label{eq:covariance}
C_{ij}&=&-m_im_j+\sum_{e\in A(G_0)}\Tr(B(e)\rho_\infty B(e)^*)(\hat\theta(e))_i(\hat\theta(e))_j\nonumber\\ 
&&+2\sum_{e\in A(G_0)}\mathrm{Tr}(B(e)\rho_\infty B(e)^*L_{{\bf e}_i})(\hat\theta(e))_j-2m_i\mathrm{Tr}(\rho_\infty L_{{\bf e}_j}).
\end{eqnarray} 
\end{thm}
\begin{rem}
Recall that $\{{\bf e}_i\}$ is the canonical basis of $\mathbb R^d$ and $L_i=L_{\hat\theta_i}$. Since ${\bf e}_i=\sum_{j=1}^d\Theta_{ij}\hat\theta_j$ (see \eqref{eq:transformation}), we can compute $L_{{\bf e}_i}$ by using $L_j$'s:
\[
L_{{\bf e}_i}=\sum_{j=1}^d{\Theta}_{ij}L_j.
\]
In the real problems, it is generally easier to compute $L_i$'s than $L_{{\bf e}_i}$'s.
\end{rem}
For the proof of the above theorem, it turns out that the methods are exactly the same as in \cite{AG-PS}. We only have a different graph structure from integer lattices and need only to modify so that it is suitable for the new structure. For the readers' convenience, however, we present the full proof in the Appendix \ref{sec:proof_clt}. 
  
\section{Examples: Hexagonal lattice}\label{sec:examples}

In this section we provide with some examples. We will consider the OQRWs on the hexagonal lattice. Look at the hexagonal lattice in Figure \ref{fig:hexagonal_lattice}.

\subsection{Preparation}\label{subsec:preparation}

We let $V_0=\{u,v\}$ and let $\{e_i\}_{i=1,2,3}$ be the three edges in $G_0$ with $\mathrm{o}(e_i)=u$ and $\mathrm{t}(e_i)=v$. (See Figure \ref{fig:hexagonal_lattice}.) The reversed edges are $\ol{e}_i$, $i=1,2,3$. We let
\[
\hat\theta(e_1)=\frac1{\sqrt{2}}(1,1),\quad\hat\theta(e_2)=\frac1{\sqrt{2}}(-1,1),\quad\hat\theta(e_3)=0,
\]
and $\hat\theta(\ol e_i)=-\hat\theta(e_i)$, $i=1,2,3$. In order to define the operators $B(e)$, $e\in A(G_0)$, let $\mathcal H_u=\mathcal H_v=\mathbb C^3$, and $\mathfrak h=\mathcal H_u\oplus \mathcal H_v\simeq \mathbb C^6$. Let $U=\left[\begin{matrix}{\bf u}_1&{\bf u}_2&{\bf u}_3\end{matrix}\right]$ and $V=\left[\begin{matrix}{\bf v}_1&{\bf v}_2&{\bf v}_3\end{matrix}\right]$ be $3\times 3$ unitary matrices with column vectors ${\bf u}_i= [u_{1i}, u_{2i},u_{3i}]^T$ and 
${\bf v}_i= [ v_{1i},v_{2i},v_{3i} ]^T$, $i=1,2,3$. For $i=1,2,3$, let $U_i$ be a $3\times 3$ matrix whose $i$th column is ${\bf u}_i$ and remaining columns are zeros. Similarly, let $V_i$ be the $3\times 3$ matrix, whose $i$th column is the vector ${\bf v}_i$ and other columns are zeros. For $i=1,2,3$, let $\tilde U_i$ and $\tilde V_i$ be $6\times 6$ matrices whose block matrices are given as follows:
\[
\tilde U_i=\left[\begin{matrix}0&0\\U_i&0\end{matrix}\right], \quad 
\tilde V_i=\left[\begin{matrix}0&V_i\\0&0\end{matrix}\right].
\]
Now we define 
\[
B(e_i):=\tilde U_i,\quad \text{and}\quad B(\ol e_i):=\tilde V_i,\quad i=1,2,3.
\]
It is easy to check that a state $\rho=\rho_u\oplus \rho_v\in \mathcal E(\mathfrak h)$ is an invariant state to the equation $\mathcal L(\rho)=\rho$, where $\mathcal L(\rho)$ is defined in  \eqref{eq:cp_equation}, if and only if it holds that 
\begin{eqnarray}
\rho_u=\sum_{i=1}^3V_i\rho_v V_i^*,\label{eq:cp_1}\\
\rho_v=\sum_{i=1}^3U_i\rho_u U_i^*.\label{eq:cp_2}
\end{eqnarray}
Consider the following (doubly) stochastic matrices.
\begin{equation}\label{eq:stochastic_matrices}
P_u:=\left[\begin{matrix}|u_{11}|^2&|u_{21}|^2&|u_{31}|^2\\
|u_{12}|^2&|u_{22}|^2&|u_{32}|^2\\|u_{13}|^2&|u_{23}|^2&|u_{33}|^2\end{matrix}\right],\quad P_v:=\left[\begin{matrix}|v_{11}|^2&|v_{21}|^2&|v_{31}|^2\\
|v_{12}|^2&|v_{22}|^2&|v_{32}|^2\\|v_{13}|^2&|v_{23}|^2&|v_{33}|^2\end{matrix}\right].
\end{equation}
\begin{prop}\label{prop:uniqueness}
If the stochastic matrices $P_uP_v$ and $P_vP_u$ are irreducible, then the equation $\mathcal L(\rho)=\rho$ has a unique solution   $\rho=\rho_u\oplus \rho_v$ with $\rho_u=\rho_v=\frac16I$. Conversely, suppose that $P_uP_v$ and $P_vP_u$ are reducible such that the corresponding Markov chains have a common decomposition into communicating classes. Then, the equation $\mathcal L(\rho)=\rho$ has infinitely many different solutions. 
\end{prop}
\Proof 
Since $U_i^*U_j=\d_{ij}P_i$, where $P_i$ is the projection onto $i$th component, by multiplying $U^*$ in the left and $U$ in the right to both terms in the equation \eqref{eq:cp_2} we get
\begin{equation}\label{eq:cp_3}
U^*\rho_vU=\mathrm{diag}\left((\rho_u)_{11},(\rho_u)_{22},(\rho_u)_{33}\right),
\end{equation}
where $\mathrm{diag}(a,b,c)$ means the diagonal matrix with entries $a,\,\,b$, and $c$. By multiplying $U$ from the left and $U^*$ from the right in the equation \eqref{eq:cp_3} we get 
\begin{equation}\label{eq:cp_4}
\rho_v=U\left(\mathrm{diag}\left((\rho_u)_{11},(\rho_u)_{22},(\rho_u)_{33}\right)\right)U^*, 
\end{equation}
and similarly we have
\begin{equation}\label{eq:cp_5}
\rho_u=V\left(\mathrm{diag}\left((\rho_v)_{11},(\rho_v)_{22},(\rho_v)_{33}\right)\right)V^*. 
\end{equation}
Comparing the diagonal components in \eqref{eq:cp_4} and \eqref{eq:cp_5}, we get
\begin{equation}\label{eq:invariant_vector1}
\left[\begin{matrix}(\rho_v)_{11}&(\rho_v)_{22}&(\rho_v)_{33}\end{matrix}\right]
=\left[\begin{matrix}(\rho_u)_{11}&(\rho_u)_{22}&(\rho_u)_{33}\end{matrix}\right]P_u,
\end{equation}
and 
\begin{equation}\label{eq:invariant_vector2}
\left[\begin{matrix}(\rho_u)_{11}&(\rho_u)_{22}&(\rho_u)_{33}\end{matrix}\right]
=\left[\begin{matrix}(\rho_v)_{11}&(\rho_v)_{22}&(\rho_v)_{33}\end{matrix}\right]P_v.
\end{equation}
Inserting the equations \eqref{eq:invariant_vector1} and \eqref{eq:invariant_vector2} to each other we have
\begin{equation}\label{eq:invariant_measure1}
\left[\begin{matrix}(\rho_u)_{11}&(\rho_u)_{22}&(\rho_u)_{33}\end{matrix}\right]
=\left[\begin{matrix}(\rho_u)_{11}&(\rho_u)_{22}&(\rho_u)_{33}\end{matrix}\right]P_uP_v,
\end{equation}
and
\begin{equation}\label{eq:invariant_measure2}
\left[\begin{matrix}(\rho_v)_{11}&(\rho_v)_{22}&(\rho_v)_{33}\end{matrix}\right]
=\left[\begin{matrix}(\rho_v)_{11}&(\rho_v)_{22}&(\rho_v)_{33}\end{matrix}\right]P_vP_u.
\end{equation}
Therefore, $\left[\begin{matrix}(\rho_u)_{11}&(\rho_u)_{22}&(\rho_u)_{33}\end{matrix}\right]$ is a stationary vector for the stochastic matrix $P_uP_v$, and $\left[\begin{matrix}(\rho_v)_{11}&(\rho_v)_{22}&(\rho_v)_{33}\end{matrix}\right]$ is a stationary vector for the stochastic matrix $P_vP_u$. 

Suppose that $P_uP_v$ and $P_vP_u$ are irreducible. Notice that since $P_uP_v$ and $P_vP_u$ are doubly stochastic matrices the uniform distribution is always a stationary distribution both for $P_uP_v$ and $P_vP_u$. Since the uniform distribution has full support, it follows that the three points (states) are all positive recurrent for the Markov chains. Now the Markov chains are irreducible, the irreducible and positive recurrent Markov chains with stochastic matrices $P_uP_v$ and $P_vP_u$ have a unique stationary state, which is, we know, the uniform distribution. Therefore, we have  
\begin{equation}\label{eq:diagonal_part}
\mathrm{diag}\left((\rho_u)_{11},(\rho_u)_{22},(\rho_u)_{33}\right)=c_uI\quad \text{and}\quad \mathrm{diag}\left((\rho_v)_{11},(\rho_v)_{22},(\rho_v)_{33}\right)=c_vI,
\end{equation}
where $c_u$ and $c_v$ are positive constants satisfying $c_u+c_v=1/3$.
We insert \eqref{eq:diagonal_part} into \eqref{eq:cp_4} and \eqref{eq:cp_5} to conclude that $\rho_u$ and $\rho_v$ are actually diagonal matrices $\frac16I$. 

Now suppose that $P_uP_v$ and $P_vP_u$ are reducible with a common decomposition of the state space, say $\{1,2,3\}$, into communicating classes. Without loss of generality, we may assume that $\{\{1,2\},\{3\}\}$ is a common communicating classes and thus $P_uP_v$ and $P_vP_u$ have the matrix forms:
\begin{equation}\label{eq:matrix_form}
P_uP_v=\left[\begin{matrix}*&*&0\\ *&*&0\\0&0&1\end{matrix}\right],\quad
P_vP_u=\left[\begin{matrix}\star&\star&0\\ \star&\star&0\\0&0&1\end{matrix}\right]. 
\end{equation}
In this case, we will show in Appendix \ref{subsec:unitary_forms} that $U$ and $V$ are of the following forms:
\begin{equation}\label{eq:unitary_forms}
U=\left[\begin{matrix} u_{11}&u_{12}&0\\
u_{21}&u_{22}&0\\0&0&u_{33}\end{matrix}\right],\quad V=\left[\begin{matrix} v_{11}&v_{12}&0\\
v_{21}&v_{22}&0\\0&0&v_{33}\end{matrix}\right].
\end{equation}
Let us then show that for any $\l\in [0,1]$, $\rho^{(\l)}=\rho_u^{(\l)}\oplus \rho_v^{(\l)}$ with $\rho_u^{(\l)}=\rho_v^{(\l)}=\frac12\mathrm{diag}(\l/2,\l/2,(1-\l))$ are all solutions to the equation $\mathcal L(\rho)=\rho$, that is, they satisfy the equations \eqref{eq:cp_1} and \eqref{eq:cp_2}. First notice that 
\[
\sum_{i=1}^3U_iU_i^*=I\quad\text{and}\quad \sum_{i=1}^3V_iV_i^*=I.
\]
In fact, if $i\neq j$, then we directly compute to see that $U_iU_j^*=0$ and $V_iV_j^*=0$. Therefore, 
\[
\sum_{i=1}^3U_iU_i^*=\left(\sum_{i=1}^3U_i\right)\left(\sum_{i=1}^3U_i^*\right)=UU^*=I,
\]
and similarly we show the second equation. We rewrite
\[
\rho_u^{(\l)}=\frac{\l}4I+\frac{2-3\l}4\left[\begin{matrix}0&0&0\\
0&0&0\\0&0&1\end{matrix}\right].
\]
Then, by the above observation,
\begin{eqnarray*}
\sum_iU_i\rho_u^{(\l)}U_i^*&=&\frac{\l}4I+\frac{2-3\l}4\sum_iU_i\left[\begin{matrix}0&0&0\\
0&0&0\\0&0&1\end{matrix}\right]U_i^*\\
&=&\frac{\l}4I+\frac{2-3\l}4|u_{33}|^2\left[\begin{matrix}0&0&0\\
0&0&0\\0&0&1\end{matrix}\right]=\rho_v^{(\l)}. 
\end{eqnarray*}
Here we have used the fact that $|u_{33}|^2=1$ from the form of unitary $U$ in \eqref{eq:unitary_forms}. 
Similarly we can show that the equation $\rho_u^{(\l)}=\sum_iV_i\rho_v^{(\l)}V_i^*$ holds.
This completes the proof.    
\EndProof
\begin{ex}\label{ex:hexagonal}
Let us consider the following two unitary matrices.
\begin{equation}\label{eq:unitary_matrices}
U_G:=\frac13\left[\begin{matrix}-1&2&2\\2&-1&2\\2&2&-1\end{matrix}\right],\quad 
U_H:=\left[\begin{matrix}\frac{1}{\sqrt{2}}&-\frac{1}{\sqrt{2}}&0\\\frac{1}{\sqrt{2}}&\frac{1}{\sqrt{2}}&0\\0&0&1\end{matrix}\right].
\end{equation}
For the choices of $(U,V)$ we consider three cases.\\[1ex]
(i) $(U,V)=(U_G,U_G)$. In this case we have 
\[
P_uP_v=P_vP_u=\frac1{81}\left[\begin{matrix}33&24&24\\24&33&24\\24&24&33
\end{matrix}\right].
\]
Thus $P_uP_v=P_vP_u$ are irreducible and we have a unique invariant state $\rho=\rho_u\oplus \rho_v$ with $\rho_u=\rho_v=\frac16I$ for the equation $\mathcal L(\rho)=\rho$.\\[1ex]
(ii) $(U,V)=(U_G,U_H)$. In this case we have 
\[
P_uP_v=P_vP_u=\frac1{18}\left[\begin{matrix}5&5&8\\
5&5&8\\8&8&2\end{matrix}\right].
\]
Thus again $P_uP_v=P_vP_u$ are irreducible and there is a unique invariant state $\rho=\rho_u\oplus \rho_v$ with $\rho_u=\rho_v=\frac16I$.\\[1ex]
(iii) $(U,V)=(U_H,U_H)$. In this case we have 
\[
P_uP_v=P_vP_u=\frac12\left[\begin{matrix}1&1&0\\1&1&0\\0&0&2
\end{matrix}\right].
\]
Here the stochastic matrix $P_uP_v=P_vP_u$ is not irreducible and the equation $\mathcal L(\rho)=\rho$ has many different solutions. We can check that for any $\l\in [0,1]$, $\rho^{(\l)}=\rho_u^{(\l)}\oplus \rho_v^{(\l)}$ with $\rho_u^{(\l)}=\rho_v^{(\l)}=\frac12\mathrm{diag}(\l/2,\l/2,(1-\l))$  are all invariant states.
\end{ex}

\subsection{Example: nonzero covariance}\label{subsec:hexagonal_nonzero_covariance}
 
From now on let us focus on a fixed model by taking $U=V=U_G$ with $U_G$ in \eqref{eq:unitary_matrices}. We want to see the mean $m$ and covariance matrix $\Sigma$ in Theorem \ref{thm:clt}. Since the unique invariant state to the equation $\mathcal L(\rho)=\rho$ is $\rho_\infty=\frac16I$, from the equation \eqref{eq:mean} it is easy to see that $m=0$. By directly computing from \eqref{eq:operator_equation}, we see that, up to a sum of a constant multiple of identity,
\[
L_1=L_{1,u}\oplus L_{1,v}, \,\,\text{with }L_{1,u}=-L_{1,v}=\frac16\left[\begin{matrix}7&0&0\\0&-2&0\\0&0&-2
\end{matrix}\right], 
\]
and
\[
L_2=L_{2,u}\oplus L_{2,v}, \,\,\text{with }L_{2,u}=-L_{2,v}=\frac16\left[\begin{matrix}-2&0&0\\0&7&0\\0&0&-2
\end{matrix}\right]. 
\]
Notice that 
\[
\Theta=\frac1{\sqrt{2}}\left[\begin{matrix}1&-1\\1&1\end{matrix}\right].
\]
Therefore, we get
\[
L_{{\bf e}_1}=\Theta_{11}L_1+\Theta_{12}L_2=L_{{\bf e}_1,u}\oplus L_{{\bf e}_1,v}, \,\,\text{with }L_{{\bf e}_1,u}=-L_{{\bf e}_1,v}=\frac3{2\sqrt{2}}\left[\begin{matrix}1&0&0\\0&-1&0\\0&0&0
\end{matrix}\right], 
\]
and
\[
L_{{\bf e}_2}=\Theta_{21}L_1+\Theta_{22}L_2=L_{{\bf e}_2,u}\oplus L_{{\bf e}_2,v}, \,\,\text{with }L_{{\bf e}_2,u}=-L_{{\bf e}_2,v}=\frac1{6\sqrt{2}}\left[\begin{matrix}5&0&0\\0&5&0\\0&0&-4
\end{matrix}\right].
\]
Now we are ready to compute the covariance matrix $\Sigma$ given in \eqref{eq:covariance}.  
Since the mean $m$ is zero, we are left with 
\begin{eqnarray}\label{eq:covariance_hexagonal}
C_{ij}&=& \sum_{e\in A(G_0)}\Tr(B(e)\rho_\infty B(e)^*)(\hat\theta(e))_i(\hat\theta(e))_j\nonumber\\ 
&&+2\sum_{e\in A(G_0)}\mathrm{Tr}(B(e)\rho_\infty B(e)^*L_{{\bf e}_i})(\hat\theta(e))_j\\
&=:&C^{(1)}_{ij}+C_{ij}^{(2)}\nonumber.
\end{eqnarray} 
For the first term, the trace part is all $1/6$ and thus we get 
\[
C^{(1)}=\frac13I.
\]
For the second term, since $\rho_\infty=\frac16I\oplus\frac16I$, we compute before taking trace,
\[
\sum_{e\in A(G_0)}(B(e)B(e)^*)(\hat\theta(e))_j=\begin{cases}\frac1{3\sqrt{2}}\left( \left[\begin{matrix}1&0&2\\0&-1&-2\\2&-2&0\end{matrix}\right]
\oplus\left[\begin{matrix}-1&0&-2\\0&1&2\\-2&2&0\end{matrix}\right]\right),&j=1,\\
\frac1{9\sqrt{2}}\left( \left[\begin{matrix}-5&4&-2\\4&-5&-2\\-2&-2&-8\end{matrix}\right]
\oplus\left[\begin{matrix}5&-4&2\\-4&5&2\\2&2&8\end{matrix}\right]
\right),&j=2.\end{cases}
\]
Using this we get 
\[
C^{(2)}=\frac19\left[\begin{matrix}3&0\\0&-1\end{matrix}\right].
\]
Thus summing those two terms we get covariance matrix
\begin{equation}\label{eq:covariance_hexagonal_grover}
\Sigma=C^{(1)}+C^{(2)}=\frac29\left[\begin{matrix}3&0\\0&1\end{matrix}\right].
\end{equation}
\begin{rem}
The movements between the points $u$ and $v$ in a single site do not contribute to the real movements. This is reflected by the fact that the variance in the vertical line ($y$-axis) is smaller than that in the horizontal line ($x$-axis) in \eqref{eq:covariance_hexagonal_grover}. 
\end{rem}
Notice that the characteristic function for the Gaussian random variable $X$ with mean zero and covariance $\Sigma$ in \eqref{eq:covariance_hexagonal_grover} is 
\begin{equation}\label{eq:characteristic_function}
\mathbb{E}(e^{i\langle {\bf t},X\rangle})=e^{-\frac19(3t_1^2+t_2^2)}.
\end{equation}

\subsection{Example: zero  covariance}\label{subsec:hexagonal_zero_covariance}
  
Let us give one more example. This example, together with the former one, we will consider again in a different view point, namely by a dual process, in the next section. 
 
For the model on the Hexagonal lattice, let us take $U=U_G$ in \eqref{eq:unitary_matrices} and $V=I$. In that case, since $P_uP_v=P_vP_u=P_u$ is irreducible, the equation $\mathcal L(\rho)=\rho$ has a unique invariant state $\rho_\infty=\frac16I\oplus\frac16I$. As before, the solutions of \eqref{eq:operator_equation} are, up to a sum of constant multiple of identity, 
\[
L_1=L_{1,u}\oplus L_{1,v}, \,\,\text{with }L_{1,u}=\left[\begin{matrix}1&0&0\\0&0&0\\0&0&0
\end{matrix}\right], \,\,L_{1,v}=0,
\]
and
\[
L_2=L_{2,u}\oplus L_{2,v}, \,\,\text{with }L_{2,u}=\left[\begin{matrix}0&0&0\\0&1&0\\0&0&0
\end{matrix}\right], \,\,L_{2,v}=0. 
\]
We then get
\[
L_{{\bf e}_1}=\Theta_{11}L_1+\Theta_{12}L_2=L_{{\bf e}_1,u}\oplus L_{{\bf e}_1,v}, \,\,\text{with }L_{{\bf e}_1,u}=\frac1{\sqrt{2}}\left[\begin{matrix}1&0&0\\0&-1&0\\0&0&0
\end{matrix}\right],\,\,L_{{\bf e}_1,v}=0, 
\]
and
\[
L_{{\bf e}_2}=\Theta_{21}L_1+\Theta_{22}L_2=L_{{\bf e}_2,u}\oplus L_{{\bf e}_2,v}, \,\,\text{with }L_{{\bf e}_2,u}=\frac1{\sqrt{2}}\left[\begin{matrix}1&0&0\\0&1&0\\0&0&0
\end{matrix}\right],\,\,L_{{\bf e}_2,v}=0.
\]
In this model, the mean and covariance matrix can be computed in the same way as before, and we get  
\begin{equation}\label{eq:covariance_hexagonal2_mean_and_covariance}
m=0,\quad \Sigma=0.
\end{equation}
This means that the measure is a Dirac measure at the origin.

\section{Dual processes}\label{sec:dual_process}

In this section we consider the dual processes for the OQRWs on the crystal lattices. The concept of dual processes was introduced in \cite{KY}, and it is an OQRW on the dual space, namely the Fourier transform space to the lattice. Since crystal lattices are intrinsically regular lattices, like the integer lattices, we can develop an analysis of Fourier transforms.  

\subsection{Fourier transform on the crystal lattices}\label{subsec:FT}

Let us denote the usual inner product in $\mathbb R^d$ by $\langle\cdot,\cdot\rangle$. The points of integer lattice $\mathbb Z^d$ and crystal lattice $\mathbb L$ are naturally embedded in $\mathbb R^d$. Recall that $\{\hat\theta_1,\cdots,\hat\theta_d\}$ is a basis for $\mathbb L$. In general they are not orthonormal. We define a one to one mapping $J:\mathbb Z^d \to \mathbb L$ by 
\begin{equation}\label{eq:embedding of integers}
J({\bf x})=\sum_{i=1}^dx_i\hat\theta_i,\quad {\bf x}=(x_1,\cdots,x_d)\in \mathbb Z^d.
\end{equation}
Embedded in $\mathbb R^d$, we see that 
\[
J({\bf x})=(\Theta^{-1})^T{\bf x},
\]
that is, 
\begin{equation}\label{eq:lattice_transform}
J=(\Theta^{-1})^T.
\end{equation}
For a function $g:\mathbb Z^d\to \mathbb C$, we also make a transformation of $g$ as a function on $\mathbb L$ by 
\begin{equation}\label{eq:function_embedding}
J(g)({x}):=g\circ J^{-1}({x}),\quad x\in \mathbb L.
\end{equation}
Let $\mathbb T:=[0,2\pi]$. Recall that for a function $g:\mathbb Z^d\to \mathbb C$, its Fourier transform is defined by 
\[
\hat g({\bf k})=\sum_{{\bf x}\in \mathbb Z^d}e^{-\langle {\bf k},{\bf x}\rangle}g({\bf x}),\quad {\bf k}\in \mathbb T^d,
\]
and its inverse Fourier transform is 
\[
g({\bf x})=\frac1{(2\p)^d}\int_{{\mathbb T}^d}e^{i\langle {\bf k},{\bf x}\rangle}\hat g({\bf k})d{\bf k}.
\]
For a function $f:\mathbb L\to \mathbb C$, we also define its Fourier transform (abusing the notations) $\hat f:\Theta({\mathbb T}^d)\to \mathbb C$ by 
\begin{eqnarray}
\hat f({\bf k})&:=&\sum_{x\in \mathbb L}e^{-i\langle {\bf k},x\rangle}f(x)\nonumber\\
&=&\sum_{{\bf x}\in \mathbb Z^d}e^{-\langle {\bf k},J{\bf x}\rangle}f\circ J({\bf x})\nonumber\\
&=&  \sum_{{\bf x}\in \mathbb Z^d}e^{-\langle \Theta^{-1} {\bf k},{\bf x}\rangle}f\circ J({\bf x}) =     \hat{f\circ J}(\Theta^{-1}{\bf k}),\quad {\bf k}\in \Theta(\mathbb T^d).\label{eq:Fourier_transform}
\end{eqnarray}
On the other hand, for $x=J({\bf x})\in \mathbb L$,
\begin{eqnarray}
f(x)&=&f\circ J(\bf x)\nonumber \\
&=&\frac1{(2\p)^d}\int_{{\mathbb T}^d}e^{i\langle {\bf k},{\bf x}\rangle}\hat {f\circ J}({\bf k})d{\bf k}\nonumber \\
&=&\frac1{(2\p)^d}\int_{{\mathbb T}^d}e^{i\langle \Theta^{-1}\Theta{\bf k},{\bf x}\rangle}\hat {f\circ J}(\Theta^{-1}\Theta{\bf k})d{\bf k}\nonumber \\
&=&\frac1{|\det\Theta|}\frac1{(2\p)^d}\int_{\Theta({\mathbb T}^d)}e^{i\langle \Theta^{-1}{\bf k},{\bf x}\rangle}\hat {f\circ J}(\Theta^{-1}{\bf k})d{\bf k}\nonumber \\
&=&\frac1{|\det\Theta|}\frac1{(2\p)^d}\int_{\Theta({\mathbb T}^d)}e^{i\langle {\bf k},{x}\rangle}\hat {f}({\bf k})d{\bf k},\quad x\in \mathbb L.\label{eq:inverse_Fourier_crystal_lattice}
\end{eqnarray}

\subsection{Dual processes}
In this subsection, we consider dual processes, which was introduced in \cite{KY}. The space $\mathcal B(\mathfrak h)$ is equipped with an inner product,
\begin{equation}\label{eq:inner_product}
\langle A,B\rangle:=\Tr(A^*B),\quad A,\,B\in \mathcal B(\mathfrak h).
\end{equation}
We let $\mathcal A:=\oplus_{x\in {\mathbb L}}\mathcal B(\mathfrak h)$ be the direct sum Hilbert space. Taking $\mathcal A_{\bf k}:=\mathcal B(\mathfrak h)$ for each ${\bf k}\in \Theta(\mathbb T^d)$, we also introduce the following direct integral Hilbert space:
\[
\hat{\mathcal A}:=\frac1{|\det\Theta|}\int_{\Theta(\mathbb T^d)}^{\oplus}\mathcal A_{\bf k}\frac1{(2\pi)^d}d{\bf k}.
\]
For each $e\in A(G_0)$, we let $T_e$ be the translation on $l^2(\mathbb L)$ defined by for each $a=(a_x)_{x\in \mathbb L}$,
\[
(T(e)a)_x=a_{x-\hat{\theta}(e)}.
\]
For any $B\in \mathcal B(\mathfrak h)$, we let $L_B$ and $R_B$ be the left and right multiplication operators, respectively, on $\mathcal B(\mathfrak h)$:
\[
L_B(A):=BA,\quad R_B(A):=AB,\quad A\in \mathcal B(\mathfrak h).
\]
Slightly abusing the notations, we also let $L_B$ and $R_B$ be the left and right multiplication operators, respectively, on $\mathcal A$ and on $\hat{\mathcal A}$: for $a=(a_x)_{x\in \mathbb L}$ and $\hat a=(a({\bf k}))_{k\in \Theta(\mathbb T)}\in\hat{\mathcal A} $,
\begin{eqnarray*}
L_B(a)&:=&(Ba_x)_{x\in \mathbb L},\quad R_B(a):=(a_xB)_{x\in \mathbb L},\\L_B(\hat a)&:=&(Ba({\bf k}))_{{\bf k}\in \Theta(\mathbb T)},\quad R_B(\hat a):=(a({\bf k})B)_{x\in \Theta(\mathbb T)}.
\end{eqnarray*}
Recall that the OQRWs on the crystal lattices are the evolution of the states of the form in \eqref{eq:states}:
\[
\rho=\sum_{x\in  \mathbb L }\left(\oplus_{u\in V_0}\rho_{(x,u)}\right)\otimes|x\rangle\langle x|.
\]
Letting $\rho_x:=\oplus_{u\in V_0}\rho_{(x,u)}\in \mathcal B(\mathfrak h) $, we regard the above state as $\rho=(\rho_x)_{x\in \mathbb L}\in \mathcal A$. Then, the dynamics of the OQRWs on the crystal lattices are represented as
\begin{equation}\label{eq:oqrw_evolution}
\rho^{(n)}=\left(\sum_{e\in A(G)}T(e)L_{B(e)}R_{B(e)^*}\right)^n\rho^{(0)}.
\end{equation}
Taking the Fourier transform, the evolution is given by 
\begin{equation}\label{eq:oqrw_evolution_Fourier_transformed}
\hat{\rho^{(n)}}({\bf k})=\left(\sum_{e\in A(G_0)}e^{-i\langle {\bf k},\hat\theta(e)\rangle}L_{B(e)}R_{B(e)^*}\right)^n\hat{\rho^{(0)}}({\bf k}),\quad {\bf k}\in \Theta(\mathbb T^d).
\end{equation}
As in \cite{KY}, we define the dual process as the process $(Y_n({\bf k}))_{{\bf k}\in \Theta(\mathbb T^d)}\in \hat {\mathcal A}$ given by 
\begin{equation}\label{eq:dual_process}
Y_n({\bf k}):=\left(\sum_{e\in A(G_0)}e^{-i\langle {\bf k},\hat\theta(e)\rangle}L_{B(e)^*}R_{B(e)}\right)^n(I_{\mathfrak h}).
\end{equation}
Notice that the positions of $B(e)$ and $B(e)^*$ are different in the equations \eqref{eq:oqrw_evolution_Fourier_transformed} and \eqref{eq:dual_process}. 
The usefulness of the dual process is given by the following theorem, which was observed in \cite[Theorem 2.3]{KY}. For a proof we refer to \cite{KY}. We just take a Fourier transform on the crystal lattice $\mathbb L$ introduced in the former subsection.
\begin{thm}\label{thm:density_by_Fourier_transform}
The probability distribution of the OQRW at time $n$ is given by 
\[
p_x^{(n)}=\frac1{|\det\Theta|}\frac1{(2\p)^d}\int_{\Theta(\mathbb T^d)}
e^{i\langle {\bf k},x\rangle}\mathrm{Tr}\left(\hat{\rho^{(0)}}({\bf k})Y_n({\bf k})\right)d{\bf k}, \quad x\in \mathbb L.
\]
\end{thm} 
That is, the Fourier transform of $(p_x^{(n)})_{x\in \mathbb L}$ is 
\[
\hat{p_\cdot^{(n)}}({\bf k})=\mathrm{Tr}\left(\hat{\rho^{(0)}}({\bf k})Y_n({\bf k})\right),\quad {\bf k}\in \Theta(\mathbb T^d).
\]
\begin{ex}
Let us consider the OQRW on the Hexagonal lattice introduced in subsection \ref{subsec:hexagonal_zero_covariance}. In this case 
\begin{equation}\label{eq:stochastic_matrix_Grover}
P_uP_v=P_vP_u=P_u=: P=\frac19\left[\begin{matrix}1&4&4\\4&1&4\\4&4&1\end{matrix}\right]
\end{equation}
is irreducible, and so by Proposition \ref{prop:uniqueness} the equation $\mathcal L(\rho)=\rho$ has a unique solution and Theorem \ref{thm:clt} applies. Here we have $\hat\theta_1=1/\sqrt{2}[1,\,1]^T$, $\hat\theta_2=1/\sqrt{2}[-1,\,1]^T$,  $\Theta=\frac1{\sqrt{2}}\left[\begin{matrix}1&-1\\1&1\end{matrix}\right]$, and hence $\det\Theta=1$. Let us define diagonal matrices 
\begin{equation}\label{eq:Fourier_diagonal}
D({\bf k}):=\mathrm{diag}(e^{-i\langle {\bf k},\hat\theta_1\rangle},e^{-i\langle {\bf k},\hat\theta_2\rangle},1),\quad {\bf k}\in \Theta(\mathbb T^2).
\end{equation}
It is promptly computed that
\begin{eqnarray*}
&&Y_n({\bf k})=A_n({\bf k})\oplus B_n({\bf k});\\
&&A_n({\bf k})=\mathrm{diag}(a_{n,1}({\bf k}),a_{n,2}({\bf k}),a_{n,3}({\bf k})),\quad  B_n({\bf k})=\mathrm{diag}(b_{n,1}({\bf k}),b_{n,2}({\bf k}),b_{n,3}({\bf k})),
\end{eqnarray*}
where the components satisfy the following recurrence relations.
\begin{equation}\label{eq:recurrence_relations}
\left[\begin{matrix}a_{n,1}({\bf k})\\a_{n,2}({\bf k})\\a_{n,3}({\bf k})\end{matrix}\right]=D({\bf k})P\left[\begin{matrix}b_{n-1,1}({\bf k})\\b_{n-1,2}({\bf k})\\b_{n-1,3}({\bf k})\end{matrix}\right],\quad 
\left[\begin{matrix}b_{n,1}({\bf k})\\b_{n,2}({\bf k})\\b_{n,3}({\bf k})\end{matrix}\right]=D({\bf k})^*\left[\begin{matrix}a_{n-1,1}({\bf k})\\a_{n-1,2}({\bf k})\\a_{n-1,3}({\bf k})\end{matrix}\right].
\end{equation}
Solving the equations \eqref{eq:recurrence_relations} with initial conditions $A_0({\bf k})=I$ and $B_0({\bf k})=I$, we get
\begin{equation}\label{eq:solutions_hexagonal}
\left[\begin{matrix}a_{n,1}({\bf k})\\a_{n,2}({\bf k})\\a_{n,3}({\bf k})\end{matrix}\right]=\tilde{A_n}({\bf k})\left[\begin{matrix}1\\1\\1\end{matrix}\right],\quad 
\left[\begin{matrix}b_{n,1}({\bf k})\\b_{n,2}({\bf k})\\b_{n,3}({\bf k})\end{matrix}\right]=\tilde{B_n}({\bf k})\left[\begin{matrix}1\\1\\1\end{matrix}\right].
\end{equation}
Here the matrices $\tilde{A}_n({\bf k})$ and $\tilde{B}_n({\bf k})$ are computed as
\begin{equation}\label{eq:solutions_hexagonal_2}
\tilde{A}_n({\bf k})=\begin{cases} D({\bf k})P^mD({\bf k})^*,&n=2m\\
D({\bf k})P^m,&n=2m-1\end{cases},\quad 
\tilde{B}_n({\bf k})=\begin{cases} P^m,&n=2m\\
P^{m-1}D({\bf k})^*,&n=2m-1.\end{cases}
\end{equation}
Notice that $P$ is diagonalized as
\[
P=S\left[\begin{matrix}1&0&0\\0&-\frac13&0\\0&0&-\frac13\end{matrix}\right]S^{-1},\quad 
S=\left[\begin{matrix}1&-1&-1\\1&0&1\\1&1&0\end{matrix}\right],\,\,
S^{-1}=\frac13\left[\begin{matrix}1&1&1\\-1&-1&2\\-1&2&-1\end{matrix}
\right].
\]
We thus get 
\[
\tilde{A}_{2m}({\bf k})=\frac13\left[\begin{matrix}1+2\left(-\frac13\right)^m
&\left(1-\left(-\frac13\right)^m\right)e^{i\langle {\bf k},\hat\theta_2-\hat\theta_1\rangle}&\left(1-\left(-\frac13\right)^m\right)
e^{-i\langle {\bf k},\hat\theta_1\rangle}\\
\left(1-\left(-\frac13\right)^m\right)e^{-i\langle {\bf k},\hat\theta_2-\hat\theta_1\rangle}&
1+2\left(-\frac13\right)^m&
\left(1-\left(-\frac13\right)^m\right)
e^{-i\langle {\bf k},\hat\theta_2\rangle}\\
\left(1-\left(-\frac13\right)^m\right)
e^{i\langle {\bf k},\hat\theta_1\rangle}&
\left(1-\left(-\frac13\right)^m\right)
e^{i\langle {\bf k},\hat\theta_2\rangle}&
1+2\left(-\frac13\right)^m
\end{matrix}\right],
\]
\[
\tilde{B}_{2m}({\bf k})=\frac13\left[\begin{matrix}1+2\left(-\frac13\right)^m
& 1-\left(-\frac13\right)^m & 1-\left(-\frac13\right)^m \\
 1-\left(-\frac13\right)^m &
1+2\left(-\frac13\right)^m&
 1-\left(-\frac13\right)^m \\
 1-\left(-\frac13\right)^m &
 1-\left(-\frac13\right)^m &
1+2\left(-\frac13\right)^m
\end{matrix}\right],
\]
and
\[
\tilde{A}_{2m-1}({\bf k})=\frac13\left[\begin{matrix}\left(1+2\left(-\frac13\right)^m\right)
e^{-i\langle {\bf k},\hat\theta_1\rangle}
&\left(1-\left(-\frac13\right)^m\right)e^{-i\langle {\bf k}, \hat\theta_1\rangle}&
\left(1-\left(-\frac13\right)^m\right)
e^{-i\langle {\bf k},\hat\theta_1\rangle}\\
\left(1-\left(-\frac13\right)^m\right)e^{-i\langle {\bf k},\hat\theta_2\rangle}&
\left(1+2\left(-\frac13\right)^m\right)e^{-i\langle {\bf k},\hat\theta_2\rangle}&
\left(1-\left(-\frac13\right)^m\right)
e^{-i\langle {\bf k},\hat\theta_2\rangle}\\
1-\left(-\frac13\right)^m
 &
 1-\left(-\frac13\right)^m &
1+2\left(-\frac13\right)^m
\end{matrix}\right],
\]
\[
\tilde{B}_{2m-1}({\bf k})=\frac13\left[\begin{matrix}\left(1+2\left(-\frac13\right)^{m-1}\right)
e^{i\langle {\bf k},\hat\theta_1\rangle}
&\left(1-\left(-\frac13\right)^{m-1}\right)e^{i\langle {\bf k}, \hat\theta_2\rangle}&
 1-\left(-\frac13\right)^{m-1} \\
\left(1-\left(-\frac13\right)^{m-1}\right)e^{i\langle {\bf k},\hat\theta_1\rangle}&
\left(1+2\left(-\frac13\right)^{m-1}\right)e^{i\langle {\bf k},\hat\theta_2\rangle}&
 1-\left(-\frac13\right)^{m-1} \\
\left(1-\left(-\frac13\right)^{m-1}\right)e^{i\langle {\bf k},\hat\theta_1\rangle}
 &
 \left(1-\left(-\frac13\right)^{m-1}\right)e^{i\langle {\bf k},\hat\theta_2\rangle} &
1+2\left(-\frac13\right)^{m-1}
\end{matrix}\right].
\]
Now finding out $Y_n({\bf k})$, we can compute the probability density $p_x^{(n)}$ explicitly by Theorem  \ref{thm:density_by_Fourier_transform}. Let us take 
\[
\rho^{(0)}:=\left(\frac16I\oplus\frac16I\right)\otimes|0\rangle\langle 0|.
\]
Then, by Theorem \ref{thm:density_by_Fourier_transform}, using the above computations we see that ,
\begin{eqnarray*}
\lim_{n\to \infty}\mathbb{E}\left[e^{i\langle {\bf t},\frac{X_n}{\sqrt{n}}\rangle}\right]&=&\lim_{n\to \infty}\sum_{x\in \mathbb L}e^{i\langle {\bf t},\frac{x}{\sqrt{n}}\rangle} p_x^{(n)}\\
&=&\lim_{n\to \infty} \hat{p_\cdot^{(n)}}(-\frac{\bf t}{\sqrt{n}})\\
&=&(1)_{{\bf k}\in \Theta({\mathbb T}^2)},
\end{eqnarray*}
that is, it is a constant function $1$. This means that the limit distribution of $X_n/\sqrt{n}$ is a Dirac measure at the origin. This result was shown in subsection \ref{subsec:hexagonal_zero_covariance}.
 In fact, we see from \eqref{eq:solutions_hexagonal_2} that for $\rho^{(0)}=\frac16I\oplus\frac16I\otimes|0\rangle\langle 0|$ and $n=2m$,
 \begin{eqnarray*}
\mathrm{Tr}(\hat{\rho^{(0)}}({\bf k})Y_n({\bf k}))&=&
 \left(\frac23+\frac13\left(-\frac13\right)^m\right)+\frac1{18}
 \left(1-\left(-\frac13\right)^m\right)\Big(e^{i\langle {\bf k},\hat\theta_1\rangle}+e^{-i\langle {\bf k},\hat\theta_1\rangle}\\
 &&\hskip 1 true cm +e^{i\langle {\bf k},\hat\theta_2\rangle}+e^{-i\langle {\bf k},\hat\theta_2\rangle}+e^{i\langle {\bf k},\hat\theta_2-\hat\theta_1\rangle}+e^{-i\langle {\bf k},\hat\theta_2-\hat\theta_1\rangle}\Big), 
 \end{eqnarray*}
and similarly for $n=2m-1$. By Theorem \ref{thm:density_by_Fourier_transform}, this means that the OQRW in this model is localized in the nearby points from the origin, the starting point. Therefore, it is obvious that we have a Dirac measure for the central limit theorem. 
\end{ex}
Next we revisit the example in subsection \ref{subsec:hexagonal_nonzero_covariance}, where the covariance matrix was nontrivial.
\begin{ex}\label{eq:hexagonal_nonzero_covariance}
We consider the OQRW on the Hexagonal lattice with $U=V=U_G$ in subsection \ref{subsec:hexagonal_nonzero_covariance}. Recall the diagonal matrices $D(\bf k)$ in \eqref{eq:Fourier_diagonal} and the stochastic matrix $P$ in \eqref{eq:stochastic_matrix_Grover}. Like in the former example, we see that 
\begin{eqnarray}\label{eq:dual_process_Grover}
&&Y_n({\bf k})=A_n({\bf k})\oplus B_n({\bf k});\\
&&A_n({\bf k})=\mathrm{diag}(a_{n,1}({\bf k}),a_{n,2}({\bf k}),a_{n,3}({\bf k})),\quad  B_n({\bf k})=\mathrm{diag}(b_{n,1}({\bf k}),b_{n,2}({\bf k}),b_{n,3}({\bf k})),\nonumber
\end{eqnarray}
where the components satisfy the following recurrence relations.
\begin{equation}\label{eq:recurrence_relations_Grover}
\left[\begin{matrix}a_{n,1}({\bf k})\\a_{n,2}({\bf k})\\a_{n,3}({\bf k})\end{matrix}\right]=D({\bf k})P\left[\begin{matrix}b_{n-1,1}({\bf k})\\b_{n-1,2}({\bf k})\\b_{n-1,3}({\bf k})\end{matrix}\right],\quad 
\left[\begin{matrix}b_{n,1}({\bf k})\\b_{n,2}({\bf k})\\b_{n,3}({\bf k})\end{matrix}\right]=D({\bf k})^*P\left[\begin{matrix}a_{n-1,1}({\bf k})\\a_{n-1,2}({\bf k})\\a_{n-1,3}({\bf k})\end{matrix}\right].
\end{equation}
In order to solve the recurrence relation, let us define
\[
D({\bf k})^{1/2}:=\mathrm{diag}(e^{-i\langle {\bf k},\hat\theta_1\rangle/2},e^{-i\langle {\bf k},\hat\theta_2\rangle/2},1),
\]
so that $(D({\bf k})^{1/2})^2=D({\bf k})$. Solving the equations \eqref{eq:recurrence_relations_Grover} with initial conditions $A_0({\bf k})=I$ and $B_0({\bf k})=I$, we get
\begin{equation}\label{eq:solutions_hexagonal_Grover}
\left[\begin{matrix}a_{n,1}({\bf k})\\a_{n,2}({\bf k})\\a_{n,3}({\bf k})\end{matrix}\right]=\tilde{A_n}({\bf k})\left[\begin{matrix}1\\1\\1\end{matrix}\right],\quad 
\left[\begin{matrix}b_{n,1}({\bf k})\\b_{n,2}({\bf k})\\b_{n,3}({\bf k})\end{matrix}\right]=\tilde{B_n}({\bf k})\left[\begin{matrix}1\\1\\1\end{matrix}\right].
\end{equation}
Here the matrices $\tilde{A}_n({\bf k})$ and $\tilde{B}_n({\bf k})$ are given by (putting $D({\bf k})=:D$, for simplicity)
\begin{eqnarray}
\tilde{A}_n({\bf k})&=&\begin{cases} D^{1/2}\left(D^{1/2}PD^*PD^{1/2}\right)^mD^{1/2},&n=2m+1,\\
D^{1/2}\left(D^{1/2}PD^*PD^{1/2}\right)^{m-1}D^{1/2}PD^*,
&n=2m,\end{cases}\label{eq:solutions_hexagonal_2_Grover_u}\\ 
\tilde{B}_n({\bf k})&=&\begin{cases} (D^*)^{1/2}\left((D^*)^{1/2}PDP(D^*)^{1/2}\right)^m(D^*)^{1/2},&n=2m+1,\\
(D^*)^{1/2}\left((D^*)^{1/2}PDP(D^*)^{1/2}\right)^{m-1}(D^*)^{1/2}PD,&n=2m.\end{cases}\label{eq:solutions_hexagonal_2_Grover_v}
\end{eqnarray}
Let us take the initial state $\rho^{(0)}=\left(\frac16I\oplus\frac16I\right)\otimes|0\rangle\langle 0|$. We then get $\hat{\rho^{(0)}}({\bf k})=\frac16I\oplus\frac16I$. We want to get the limit
\begin{eqnarray}\label{eq:characteristic_Grover}
\lim_{n\to \infty}\mathbb{E}(e^{i\langle {\bf t},\frac{X_n}{\sqrt{n}}\rangle})&=&\lim_{n\to \infty}\hat{p_\cdot^{(n)}}(-\frac{{\bf t}}{\sqrt{n}})\nonumber\\
&=&\frac16\lim_{n\to \infty}\mathrm{Tr}\left(Y_n(-\frac{{\bf t}}{\sqrt{n}})\right).
\end{eqnarray}
Using \eqref{eq:recurrence_relations_Grover} - \eqref{eq:solutions_hexagonal_2_Grover_v}, we can find the limit in \eqref{eq:characteristic_Grover}. One may get a help from Mathematica to get the limit, but an analytic proof of this is given in Appendix \ref{sec:analytic proof}. Anyway, the limit is as follows: 
\begin{equation}\label{eq:limit_characteristic_function}
\lim_{n\to \infty}\mathbb{E}(e^{i\langle {\bf t},\frac{X_n}{\sqrt{n}}\rangle})=e^{-\frac19(3t_1^2+t_2^2)}.
\end{equation}
Notice that this is the same as that obtained in \eqref{eq:characteristic_function}, subsection \ref{subsec:hexagonal_nonzero_covariance}. That is, the process $X_n/\sqrt{n}$ converges in distribution to a Gaussian measure with mean zero and covariance $\Sigma$ in \eqref{eq:covariance_hexagonal_grover}.  
\end{ex}
                                                                                                                                    
\appendix
\section{Proof of CLT}\label{sec:proof_clt}
            
For any $l\in \mathbb R^d$, we have 
\begin{eqnarray*}
X_n\cdot l&=&X_0\cdot l+\sum_{k=1}^n(X_k-X_{k-1})\cdot l\\
&=&X_0\cdot l+\sum_{k=1}^n\hat\theta(Y_k)\cdot l\\
&=&X_0\cdot l+\sum_{k=1}^n\big((I-P)f(\rho_k,Y_k)+m\cdot l\big).
\end{eqnarray*}
Therefore, 
\begin{eqnarray*}
X_n\cdot l- n(m\cdot l)&=&\sum_{k=2}^n\left(f(\rho_k,Y_k)-Pf(\rho_{k-1},Y_{k-1})\right)\\
&&+X_0\cdot l+f(\rho_1,Y_1)-Pf(\rho_n,Y_n)\\
&=:&M_n+R_n,
\end{eqnarray*}
with
\begin{eqnarray*}
M_n&=&\sum_{k=2}^n\left(f(\rho_k,Y_k)-Pf(\rho_{k-1},Y_{k-1})\right)\\
R_n&=&X_0\cdot l+f(\rho_1,Y_1)-Pf(\rho_n,Y_n).
\end{eqnarray*}
Clearly $(M_n)_{n\ge 2}$ is a centered Martingale w.r.t. the filtration $(\mathcal F_n)_{n\ge 2}$ where $\mathcal F_n:=\s\{(\rho_k,X_k):k\le n\}$. $\{R_n\}$ is a bounded sequence as the following lemma shows.
\begin{lem}\label{lem:bounded_remainder}
The sequence $(|R_n|)_{n\in \mathbb N}$ is uniformly bounded.
\end{lem}
\Proof
By definition
\[
Pf(\rho_n,Y_n)=\Tr(\rho_n L_l)+ m \cdot l.
\]
We notice that $|\Tr(\rho_n L_l )|\le  \|L_l\|$ uniformly for $n$. This completes the proof.
\EndProof\\
We use here the same central limit theorem introduced in \cite[Theorem 5.4]{AG-PS} (see also the reference therein).
\begin{thm}\label{thm:clt_general}
Let $(M_n)_{n\in \mathbb N}$ be a centered, square integrable, real martingale for the filtration $(\mathcal F_n)_{n\in \mathbb N}$. If 
\begin{equation}\label{eq:clt_1}
\lim_{n\to \infty}\frac1n\sum_{k=1}^n\mathbb E\left[(\D M_k)^21_{|\D M_k|\ge \e \sqrt{n}}|\mathcal F_{k-1}\right]=0
\end{equation}
and 
\begin{equation}\label{eq:clt_2}
\lim_{n\to \infty}\frac1n\sum_{k=1}^n\mathbb E\left[(\D M_k)^2|\mathcal F_{k-1}\right]=\s^2
\end{equation}
for some $\s\ge 0$, then $M_n/\sqrt{n}$ converges in distribution to a $N(0,\s^2)$ distribution.
\end{thm}
We compute 
\begin{eqnarray*}
\D M_k&=&f(\rho_k,Y_k)-Pf(\rho_{k-1},Y_{k-1})\\
&=&\Tr(\rho_kL_l )-\Tr(\rho_{k-1}L_l )+(\hat\theta(Y_k)-m )\cdot l.
\end{eqnarray*}
Therefore, we get
\[
|\D M_k|\le 2 \|L_l \|+ \|m \|\,\|l\|+\max_{e\in A(G_0)}\|\hat\theta(e)\|\,\|l\|.
\]
The condition \eqref{eq:clt_1} obviously holds. Next we show the condition \eqref{eq:clt_2}. We see that
\begin{eqnarray*}
(\D M_k)^2&=&\Tr(\rho_kL_l )^2- \Tr(\rho_{k-1}L_l )^2\\
&&-2\Tr(\rho_{k-1}L_l)\D M_k\\
&&+(\hat\theta(Y_k)\cdot l-m \cdot l)^2+2\Tr(\rho_kL_l )(\hat\theta(Y_k)\cdot l-m \cdot l)\\
&=:&T_1+T_2+T_3,
\end{eqnarray*}
where $T_i$, $i=1,2,3$, are respectively the quantities in the lines. The term $\mathbb{E}[T_1|\mathcal F_{k-1}]$ is equal to 
\[
\mathbb {E}[\Tr(\rho_kL_l )^2|\mathcal F_{k-1}]-\Tr(\rho_kL_l )^2+\Tr(\rho_kL_l )^2-\Tr(\rho_{k-1}L_l )^2.
\]
The term $\mathbb {E}[\Tr(\rho_kL_l )^2|\mathcal F_{k-1}]-\Tr(\rho_kL_l )^2$ is the increment of a martingale, say $(Z_n)_{n\in \mathbb N}$, and it is bounded independently of $k$. Hence $Z_n/n$ converges almost surely to 0. 
The term $\Tr(\rho_kL_l )^2-\Tr(\rho_{k-1}L_l )^2$, when summed up to $n$ gives  $\Tr(\rho_nL_l )^2-\Tr(\rho_{1}L_l )^2$ and hence converges to 0 when divided by $n$.

The term $\mathbb{E}[T_2|\mathcal F_{k-1}]$ clearly vanishes:
\[
\mathbb{E}[\Tr(\rho_{k-1}L_l )\D M_k|\mathcal F_{k-1}]=\Tr(\rho_{k-1}L_l )\mathbb{E}[\D M_k|\mathcal F_{k-1}]=0.
\] 
Finally we compute $\mathbb{E}[T_3|\mathcal F_{k-1}]$.
\begin{eqnarray*}
\mathbb{E}[T_3|\mathcal F_{k-1}]&=&\mathbb{E}\big[(\hat\theta(Y_k)\cdot l)^2-2(m \cdot l)(\hat\theta(Y_k)\cdot l)+(m \cdot l)^2\\
&&+2\Tr\big(\rho_kL_l (\hat\theta(Y_k)\cdot l-m \cdot l)\big)|\mathcal F_{k-1}\big]\\
&=&\sum_{ e\in A(G_0) }\Tr\big(B(e)\rho_{k-1}B(e)^*\big) \big[(\hat\theta(e)\cdot l)^2-2(m \cdot l)(\hat\theta(e)\cdot l)\big]+(m \cdot l)^2\\
&&+2\sum_{ e\in A(G_0) }\Tr\big(B(e)\rho_{k-1}B(e)^*L_l \big) \big[\hat\theta(e)\cdot l-m \cdot l\big]\\
&=&\Tr \Big(\rho_{k-1} \Gamma_l \Big),
\end{eqnarray*}
where for $l\in \mathbb R^d$, $\Gamma_l$ is defined by
\[
\Gamma_l :=\sum_{ e\in A(G_0) }\big [B(e)^*B(e)\big(\hat\theta(e)\cdot l-m \cdot l\big)^2 +2B(e)^*L_l B(e)\big(\hat\theta(e)\cdot l-m \cdot l\big)\big].
\]
Now by the above observations and the ergodicity property introduced in Remark \ref{rem:ergodicity}, and the hypothesis (H), we see that
\[
\frac1n\sum_{k=3}^n\mathbb E\big[(\D M_k)^2|\mathcal F_{k-1}\big]
\]
converges almost surely to 
\[
\s_l^2=\Tr(\rho_\infty\Gamma_l).
\]
In order to get the covariance matrix, we compute $\s_l^2$. By using the fact that $\mathcal L$ leaves $\rho_\infty$ invariant, it is not hard to compute
\begin{eqnarray*}
\s_l^2&=&\Tr(\rho_\infty\Gamma_l)\\
&=&-(m\cdot l)^2+\sum_{e\in A(G_0)}\Tr(B(e)\rho_\infty B(e)^*)(\hat\theta(e)\cdot l)^2\\
&&+2\sum_{e\in A(G_0)}\Tr(B(e)\rho_\infty B(e)^*L_l)(\hat\theta(e)\cdot l)-2(m\cdot l)\Tr(\rho_\infty L_l).
\end{eqnarray*}
Therefore, if we put $Z=\lim_{n\to \infty}\frac{X_n-nm}{\sqrt{n}}$, convergence in law, then $Z$ has mean zero and covariance matrix $\Sigma=(C_{ij})_{i,j=1}^d$, with
\begin{eqnarray}\label{eq:covariance_in_appendix}
C_{ij}&=&-m_im_j+\sum_{e\in A(G_0)}\Tr(B(e)\rho_\infty B(e)^*)(\hat\theta(e))_i(\hat\theta(e))_j\nonumber\\ 
&&+2\sum_{e\in A(G_0)}\Tr(B(e)\rho_\infty B(e)^*L_{{\bf e}_i})(\hat\theta(e))_j-2m_i\Tr(\rho_\infty L_{{\bf e}_j}).
\end{eqnarray} 

\section{Proof of \eqref{eq:unitary_forms}}\label{subsec:unitary_forms}
Recall the doubly stochastic matrices $P_u$ and $P_v$ in \eqref{eq:stochastic_matrices} and suppose that both of the stochastic matrices $P_uP_v$ and $P_vP_u$ have the form in \eqref{eq:matrix_form}. First we show that $v_{31}$ and $v_{32}$ can not be both nonzero. In fact, suppose that they are both nonzero. Then, from $(P_uP_v)_{13}=0$ we get 
$u_{11}=0$ and $u_{21}=0$, and therefore, $|u_{31}|=1$, since $P_u$ is a stochastic matrix. Similarly, computing $(P_uP_v)_{23}=0$, we get $u_{12}=0$, $u_{22}=0$, and hence $|u_{32}|=1$. Then, $P_u$ looks like
\[
 P_u=\left[\begin{matrix}0&0&1\\0&0&1\\1&1&?\end{matrix}\right],
 \]
and this is impossible because $P_u$ is a doubly stochastic matrix. Therefore, at least one of $v_{31}$ and $v_{32}$ is zero. Suppose that $v_{31}\neq 0$ and $v_{32}=0$. As before, we compute $(P_uP_v)_{13}=0$. 
Since $v_{31}\neq 0$, we must have $u_{11}=0$.  Similarly, computing $(P_uP_v)_{23}=0$, we get $u_{12}=0$. Therefore, using the fact that $P_u$ is a stochastic matrix, $P_u$ looks like
 \begin{equation}\label{eq:P_u_form}
 P_u=\left[\begin{matrix}0&*&*\\0&*&*\\1&0&0\end{matrix}\right].
 \end{equation}
 Now computing $(P_uP_v)_{31}=0$ and $(P_uP_v)_{32}=0$ we have $v_{11}=0$ and $v_{21}=0$. Therefore, we have 
 \[
 P_v=\left[\begin{matrix}0&0&1\\ *&*&0\\ *&*&0\end{matrix}\right].
 \]
 Then, multiplying $P_u$ and $P_v$ we get
 \[
P_u P_v=\left[\begin{matrix} *&*&0\\ *&*&0\\ 0&0&1\end{matrix}\right].
 \] 
Reversing the roles of $P_u$ and $P_v$, we must also have at least one of $u_{31}$ and $u_{32}$ is equal to zero. If $u_{31}\neq 0$ and $u_{32}=0$, from \eqref{eq:P_u_form} we conclude that $P_u$ must be 
\[
 P_u=\left[\begin{matrix}0&0&1\\0&1&0\\1&0&0\end{matrix}\right].
\]  
Then, computing $P_vP_u$ we get 
\[
P_vP_u=\left[\begin{matrix}0&0&1\\ *&*&0\\ *&*&0\end{matrix}\right]\left[\begin{matrix}0&0&1\\0&1&0\\1&0&0\end{matrix}\right]
=\left[\begin{matrix}1&0&0\\0&*&*\\0&*&*\end{matrix}\right].
\]
But this contradicts \eqref{eq:matrix_form}. In the case $u_{31}=0$ and $u_{32}\neq 0$, using again \eqref{eq:P_u_form} we have  
\[
 P_u=\left[\begin{matrix}0&1&0\\0&0&1\\1&0&0\end{matrix}\right].
\]  
Then it follows that 
\[
P_vP_u=\left[\begin{matrix}0&0&1\\ *&*&0\\ *&*&0\end{matrix}\right]\left[\begin{matrix}0&1&0\\0&0&1\\1&0&0\end{matrix}\right]
=\left[\begin{matrix}1&0&0\\0&*&*\\0&*&*\end{matrix}\right].
\]
It again contradicts \eqref{eq:matrix_form}. This shows that the case $v_{31}\neq 0$ and $v_{32}=0$ is impossible. Similar argument shows that the other case $v_{31}=0$ and $v_{32}\neq 0$ is also  impossible. Therefore, we conclude that
we must have $v_{31}=0$ and $v_{32}=0$, and $P_v$ has the form:
\[
 P_v=\left[\begin{matrix}*&*&0\\ *&*&0\\ 0&0&1\end{matrix}\right].
 \]  
Exchanging the roles of $P_u$ and $P_v$ we see that $P_u$ is also of this form. The proof is completed. 

\section{Proof of \eqref{eq:limit_characteristic_function}}\label{sec:analytic proof}

In this section we provide with an analytic proof of \eqref{eq:limit_characteristic_function}, the limit characteristic function of the scaled OQRW in Example \ref{eq:hexagonal_nonzero_covariance}. We will prove the following theorem.
\begin{thm}\label{thm:limit_characteristic_function}
For any ${\bf t}\in \mathbb R^2$,
\[
\lim_{n\to \infty}\mathbb E\left[e^{i\langle{\bf t},X_n/\sqrt{n}\rangle}\right]=e^{-\frac29\e^2({\bf t})},
\]
where 
\[
\e^2({\bf t})=\langle{\bf t},\hat\theta_1\rangle^2+\langle{\bf t},\hat\theta_2\rangle^2-\langle{\bf t},\hat\theta_1\rangle\langle{\bf t},\hat\theta_2\rangle.
\]
\end{thm}
\begin{cor}\label{cor:limit_characteristic_function}
As in the Example \ref{eq:hexagonal_nonzero_covariance}, for $\hat\theta_1=1/\sqrt{2}[1,1]^T$ and $\hat\theta_2=1/\sqrt{2}[-1,1]^T$, 
\[
\lim_{n\to \infty}\mathbb E\left[e^{i\langle{\bf t},X_n/\sqrt{n}\rangle}\right]=e^{-\frac19(3t_1^2+t_2^2)}.
\]
\end{cor}
This is a proof of \eqref{eq:limit_characteristic_function}. \\[1ex]
\Proof[ of Theorem \ref{thm:limit_characteristic_function}]
Recall that we are taking the initial state $\rho^{(0)}=\left(\frac16I\oplus\frac16I\right)\otimes|0\rangle\langle 0|$, and hence we have $\hat{\rho^{(0)}}({\bf k})=\frac16I\oplus\frac16I$. Therefore, the Fourier transform of the probability distribution of OQRW at time $n$ is given by (see Theorem \ref{thm:density_by_Fourier_transform})
\begin{eqnarray*}
\hat{p_\cdot^{(n)}}({\bf k})&=&\mathrm{Tr}\left(\hat{\rho^{(0)}}({\bf k})Y_n({\bf k})\right) \\
&=&\mathrm{Re}\langle u_0,\tilde A_n({\bf k})u_0\rangle.
\end{eqnarray*}
Here $u_0:=1/\sqrt{3}[1,1,1]^T$ and we have used the fact that 
$\tilde B_n({\bf k})=\overline{\tilde A_n({\bf k})}$.
Putting $\theta_j=-\langle {\bf k},\hat\theta_j\rangle$, $j=1,2$, for simplicity, we have $D=\mathrm{diag}(e^{i\theta_1},e^{i\theta_2},1)$. 
By defining $P_{\pm}:=D^{\pm1/2}PD^{\mp 1/2}$, we can write
\[
\tilde A_n=\begin{cases}D^{1/2}(P_+P_-)^mD^{1/2},&n=2m+1,\\
D^{1/2}(P_+P_-)^{m-1}P_+D^{-1/2},&n=2m.
\end{cases}
\]
From now on we only consider $n=2m+1$. The other case can be done similarly. Putting $u_{\pm}:=D^{\pm 1/2}u_0$, we have 
\begin{equation}\label{eq:computation_of_FT}
\hat{p_\cdot^{(n)}}({\bf k})=\mathrm{Re}\langle u_-,(P_+P_-)^mu_+\rangle,\quad (n=2m+1).
\end{equation}
We notice that 
\[
P_{\pm}=\frac43|u_{\pm}\rangle\langle u_{\pm}|-\frac13I.
\]
By directly computing we get
\begin{eqnarray*}
(P_+P_-)|u_+\rangle&=&\left(\left(\frac43\mu\right)^2-\frac13\right)|u_+\rangle-\frac49\langle u_-,u_+\rangle|u_-\rangle\\
(P_+P_-)|u_-\rangle &=&\frac43\langle u_+,u_-\rangle|u_+\rangle-\frac13|u_-\rangle.
\end{eqnarray*}
Here, 
\[
\mu:=|\langle u_+,u_-\rangle|=\frac13\left|1+e^{i\theta_1}+e^{i\theta_2}\right|.
\]
Therefore, we see that $(P_+P_-)$ has an invariant subspace $\mathcal L:=\mathrm{span}\{u_+,u_-\}$. Let $M$ be the matrix representation of $(P_+P_-)_{|\mathcal L}$ w.r.t. $\{u_+,u_-\}$, i.e.,
\[
M=\left[\begin{matrix}\left(\frac43\mu\right)^2-\frac13&\frac43\langle u_+,u_-\rangle\\-\frac49\langle u_-,u_+\rangle&-\frac13\end{matrix}\right].
\]
By noticing that 
\[
\langle u_-,au_++bu_-\rangle=a\langle u_-,u_+\rangle +b,
\]
we have from \eqref{eq:computation_of_FT},
\begin{equation}\label{eq:computation_of_FT2}
\hat{p_\cdot^{(n)}}({\bf k})=\mathrm{Re}\left\langle \left[\begin{matrix}\langle u_+,u_-\rangle\\1\end{matrix}\right],M^m\left[\begin{matrix}1\\0
\end{matrix}\right]\right\rangle.
\end{equation}
Let $\{\l_{\pm}\}$ and $\{v_{\pm}\}$ be the eigensystem of $M$:
\[
Mv_{\pm}=\l_{\pm} v_{\pm}.
\]
By directly computing we have
\begin{eqnarray*}
\l_{\pm}&=&\frac89\mu^2-\frac13\pm\frac49\mu\sqrt{4\mu^2-3},\\
v_{\pm}&=&\left[\begin{matrix}\frac89\mu^2\pm\frac49\mu\sqrt{4\mu^2-3}\\
-\left(\frac23\right)^2\langle u_-,u_+\rangle\end{matrix}\right].
\end{eqnarray*}
Let $V:=\left[ v_+\,\,v_-\right]$ be the similarity matrix for the diagonalization of $M$. Then from \eqref{eq:computation_of_FT2} we have  
\begin{equation}\label{eq:computation_of_FT3}
\hat{p_\cdot^{(n)}}({\bf k})=\mathrm{Re}\left\langle V^*
\left[\begin{matrix}\langle u_+,u_-\rangle\\1\end{matrix}\right],\left[\begin{matrix}\l_+^m&0\\0&\l_-^m\end{matrix}\right]V^{-1}
\left[\begin{matrix}1\\0
\end{matrix}\right]\right\rangle.
\end{equation}
By directly computing we get
\begin{eqnarray}
V^{-1}\left[\begin{matrix}1\\0
\end{matrix}\right]&=&\frac9{8\mu\sqrt{4\mu^2-3}}\left[\begin{matrix}
1\\-1\end{matrix}\right]\label{eq:V_inverse}\\
V^*\left[\begin{matrix}\langle u_+,u_-\rangle\\1\end{matrix}\right]&=&
\langle u_+,u_-\rangle\left[\begin{matrix}\frac89\mu^2+\frac49\mu
\sqrt{4\mu^2-3}\\
\frac89\mu^2-\frac49\mu
\sqrt{4\mu^2-3}\end{matrix}\right]-\left(\frac23\right)^2\langle u_+,u_-\rangle\left[\begin{matrix}1\\1\end{matrix}\right].\label{eq:V_star}
\end{eqnarray}

Now let us consider the asymptotics of $\hat{p_\cdot^{(n)}}({\bf k})$ for large $n$ replacing ${\bf k}$ by $-{\bf t}/\sqrt{n}$. In this case, recalling now $\theta_j=-\frac1{\sqrt{n}}\langle {\bf t},
\hat\theta_j\rangle$, $j=1,2$, we see that
\[
\mu=|\langle u_+,u_-\rangle|=\frac13\left|1+e^{i\theta_1}+e^{i\theta_2}\right|=1+O(1/\sqrt{n}).
\]
Therefore, by \eqref{eq:V_inverse} and \eqref{eq:V_star}, we get as $n\to \infty$,
\begin{eqnarray}
V^{-1}\left[\begin{matrix}1\\0
\end{matrix}\right]&\to& \frac98\left[\begin{matrix}1\\-1\end{matrix}\right],
\label{eq:V_inverse_limit}\\ 
V^*\left[\begin{matrix}\langle u_+,u_-\rangle\\1\end{matrix}\right]&\to&\frac89\left[\begin{matrix}1\\0
\end{matrix}\right]. 
\label{eq:V_star_limit}
\end{eqnarray}
To get the asymptotics of $\l_+^m$ and $\l_-^m$, we need to get more sharp estimate for $\mu$. Notice that
\begin{eqnarray*}
\mu^2&=&\frac19\left|1+e^{i\theta_1}+e^{i\theta_2}\right|^2=\frac19
(3+2\cos\theta_1+2\theta_2+2\cos(\theta_2-\theta_1))\\
&=&1-\frac2{9n}\e^2({\bf t})+o(1/n),
\end{eqnarray*}
where
\[
\e^2({\bf t})=\langle{\bf t},\hat\theta_1\rangle^2+\langle{\bf t},\hat\theta_2\rangle^2-\langle{\bf t},\hat\theta_1\rangle\langle{\bf t},\hat\theta_2\rangle.
\]
Thus we have 
\[
\mu=1-\frac1{9n}\e^2({\bf t})+o(1/n).
\]
Recalling $\l_+=\frac89\mu^2-\frac13+\frac49\mu\sqrt{4\mu^2-3}$, we have 
\[
\l_+=1-\frac4{9n}\e^2({\bf t})+o(1/n).
\]
Thus as $n\to \infty$,
\begin{equation}\label{eq:lambda_plus_limit}
\l_+^m=\l_+^{\frac{n-1}2}=\left(1-\frac4{9n}\e^2({\bf t})+o(1/n)\right)^{\frac{n-1}2}\to e^{-\frac29\e^2({\bf t})}.
\end{equation}
Similarly,
\[
\l_-=\frac89\mu^2-\frac13+\frac49\mu\sqrt{4\mu^2-3}=\frac19+o(1/n),
\]
and hence as $n\to \infty$,
\begin{equation}\label{eq:lambda_minus_limit}
\l_-^m\to 0.
\end{equation}
Now plugging the results \eqref{eq:V_inverse_limit}-\eqref{eq:lambda_minus_limit} into \eqref{eq:computation_of_FT3} we get as $n\to \infty$,
\[
\hat{p_\cdot^{(n)}}({\bf t}/\sqrt{n})\to e^{-\frac29\e^2({\bf t})}.
\]
This proves Theorem \ref{thm:limit_characteristic_function}.
\EndProof
 
\noindent\textbf{Acknowledgments}. E. Segawa acknowledges financial supports from the Grant-in-Aid for Young Scientists (B) and of Scientific Research
(B) Japan Society for the Promotion of Science (Grant Nos. 16K17637, 16K03939). The research by H. J. Yoo was supported by Basic Science Research Program through the National
Research Foundation of Korea (NRF) funded by the Ministry of Education (NRF-2016R1D1A1B03936006).

\end{document}